\documentclass[prb,twocolumn,superscriptaddress,showpacs,preprintnumbers,amsmath,amssymb,floats]
{revtex4}

\usepackage{txfonts}
\usepackage{amssymb}
\usepackage{graphicx}

\begin{document}

\title{Extraordinary  quasiparticle scattering and bandwidth-control by dopants in iron-based superconductors}

\author{Z. R. Ye}
\author{Y. Zhang}
\author{F. Chen}
\author{M. Xu}
\author{J. Jiang}
\author{X. H. Niu}
\author{C. H. P. Wen}
\affiliation{State Key Laboratory of Surface Physics, Department of
Physics,  and Advanced Materials Laboratory, Fudan University,
Shanghai 200433, People's Republic of China}

\author{L. Y. Xing}
\author{X. C. Wang}
\author{C. Q. Jin}
\affiliation{Institute of Physics, Chinese Academy of Sciences, Beijing 100190, China}

\author{B. P. Xie}\email{bpxie@fudan.edu.cn}
\author{D. L. Feng}\email{dlfeng@fudan.edu.cn}
\affiliation{State Key Laboratory of Surface Physics, Department of
Physics,  and Advanced Materials Laboratory, Fudan University,
Shanghai 200433, People's Republic of China}

\date{\today}

\begin{abstract}

The diversities in crystal structures and ways of doping result in  extremely  diversified phase diagrams 
for iron-based superconductors.
With angle-resolved photoemission spectroscopy (ARPES), we have systematically studied the effects of chemical substitution on the electronic structure of various series of  iron-based superconductors. In addition to the control of Fermi surface topology by heterovalent doping, we found two more extraordinary effects of doping:  1.  the site and band dependencies of quasiparticle scattering; and more importantly 2. the ubiquitous and significant bandwidth-control by both isovalent and heterovalent dopants in the iron-anion layer. Moreover, we found that the bandwidth-control could be achieved by either applying the chemical pressure or doping electrons, but not by doping holes. Together with  other findings provided here, these results complete the microscopic picture of the  electronic effects of dopants,
which facilitates a unified understanding of the diversified phase diagrams and resolutions to many open issues of various iron-based superconductors. 

\end{abstract}


\maketitle

\section{Introduction}

Chemical substitution, or doping, is the most common way to tune the properties of a correlated material. Besides introducing impurity scatterings, the dopants can affect the electronic properties of materials in two ways: 1. changing the chemical potential and Fermi surface by carrier doping, known as filling-control; 2. tuning the hopping of electrons or bandwidth, known as bandwidth-control, which could affect the relative strength of electronic interactions \cite{MIT,HCXu}.

The high temperature superconductivity in cuprates is  induced by doping a few percent of additional holes or electrons into their  \emph{insulating} antiferromagnetic parent compounds. Similarly,  
the dome-like superconducting regions in the phase diagrams  of most  iron-based high-temperature superconductors (FeHTS's)  are reached by doping their \emph{metallic} collinear-antiferromagnetic (CAF) parent compounds as well \cite{FeReview1,FeReview2}.
However intriguingly, no matter it is doped with heterovalent elements to introduce holes  [eg. Ba$_{1-x}$K$_x$Fe$_2$As$_2$ (ref.~\onlinecite{XHChenBaK})] or electrons [eg. Ba(Fe$_{1-x}$Co$_x$)$_2$As$_2$ (ref.~\onlinecite{BaCophase})], or it is doped with isovalent elements to introduce compressional [eg. BaFe$_2$(As$_{1-x}$P$_x$)$_2$ (refs.~\onlinecite{BaPphase, Zirong122})]  or tensile [eg. Ba(Fe$_{1-x}$Ru$_x$)$_2$As$_2$ (ref.~\onlinecite{BaRuphase})]  strain,  the generic features of the phase diagrams, such as a superconducting dome, are qualitatively the same.
While the superconductivity in cuprates is extremely sensitive to impurity scattering \cite{CuprateIm},  the  superconducting transition temperature, $T_C$, of FeHTS's seems to be much less sensitive against various common impurities \cite{FeImpurity}.
Taking Ba(Fe$_{1-x}$Co$_x$)$_2$As$_2$ as an example, though the cobalt (Co) dopants are in the iron (Fe) layer, the maximal $T_C$ is as high as 22~K for 8\% doping \cite{BaCoTrans}. Such robustness of $T_C$ was proposed to be important for understanding the pairing symmetry of superconductivity \cite{GapIm, GapReview}.
On the other hand,  the sizes of the superconducting domes vary significantly in various families of FeHTS's \cite{XHChenBaK, BaCophase, BaPphase, Zirong122, BaRuphase, FeTeSephase, NaCophase, LiCophase, 1111phase}, unlike the universal carrier doping range observed in cuprates \cite{Cupratephase}.
It is thus intriguing to study how the dopant affects the electronic structure in FeHTS's.

In addition to the issues related to the overall  phase diagram, there are various other unexplained doping behaviors as well. For example,
in Ba(Fe$_{1-x}$Co$_x$)$_2$As$_2$, through electron doping, the central pockets change from hole type to electron type, known as the Lifshitz transition, which was found to be accompanied with the disappearance of superconductivity \cite{BaColifshitz}. The nesting between the hole and electron pockets was also suggested to be responsible for the maximal $T_C$'s in Ba(Fe$_{1-x}$Co$_x$)$_2$As$_2$ and Ba$_{1-x}$K$_x$Fe$_2$As$_2$ (ref.~\onlinecite{HDingreview}). However, their counter examples have been raised,  and  the role of  the Fermi surface topology on superconductivity is still an open debate \cite{YZhangKFeSe, DLFeSe, XJFeSe}.
There is also an empirical relation between the highest $T_C$, of each series of FeHTS's and an optimal anion-Fe-anion bond angle or an optimal height of anion with respect to the Fe layer (referred to as anion height).   It was found that $T_C$ maximizes, when the bond angle is around $109.47^{\circ}$  or the anion height is around 1.38\AA \cite{BondAngle, BondLength}. However, so far, the direct connection among lattice, electronic structure, and $T_C$ is yet to be established.
All these unusual and seemingly unrelated puzzles request a deeper and more comprehensive understanding of the doping effects in FeHTS's.
 The diversity of the materials and diversified ways of doping add complexities to the task; however, they also provide an opportunity, because a systematical study of various series of FeHTS's would help to pin down the  common and critical ingredients of the unconventional superconductivity in these compounds.


We here present our systematic study of the doping effects on the electronic structures of the so-called 11, 111, and 122 series of FeHTS's with  angle-resolved photoemission spectroscopy (ARPES).
For the three essential consequences of doping:  filling-control, impurity scattering, and bandwidth-control in FeHTS's, our data reveal many extraordinary behaviors of the latter two consequences, which turns out  to help answer many current unresolved issues and puzzles related to the doping, and help unify the diversified phase diagrams. More specifically, we found that

\begin{enumerate}

\item  the quasiparticle scattering induced by the dopants exhibits a band-selective and site-dependent behavior. All  the  bands, except  the  $d_{xy}$ hole-like band around the zone center,  are inert to the doped impurities. Moreover, the scattering strength of the  $d_{xy}$ hole-like band depends on the site of the dopants. The dopants at Fe site cause the strongest scattering, and those at the anion site cause sizable scattering, while those off the Fe-anion plane do not cause much scattering.

\item  both the heterovalent doping and isovalent doping cause dramatic change to the bandwidth.  Remarkably, the Co dopants at Fe site cause the strongest bandwidth enhancement, and phosphorus (P) or tellurium (Te) dopants at the anion site increase the bandwidth moderately, while the potassium (K) dopants off the Fe-anion plane do not affect the bandwidth noticeably. We found that the chemical pressure, such as the change of bond length, plays an important role on the bandwidth-control. Meanwhile, the carrier doping affects the bandwidth in a particle-hole asymmetric fashion, which highlights the distinctive nature of electronic correlations in FeHTS's.

\item  the Fermi surface topology in FeHTS's shows a large diversity. We further demonstrate that  the disappearance of  certain hole pockets does not have to correspond to a diminishing $T_C$.  Moreover, for the heavily electron-doped compounds with the same Fermi surface topology, only systems with narrow bandwidths exhibit superconductivity.

\end{enumerate}

Many previous studies have tried to establish the relationship between $T_C$ and the Fermi surface topology, mainly focusing on the filling-control aspect of the doping.  However, many of such attempts, such as the Fermi surface nesting scenario for optimizing $T_C$, have been proven just accidental in some peculiar compounds \cite{BorisenkoLiFeAs, YZhangKFeSe, DLFeSe, XJFeSe}.
In the present paper,  we  further  point out that  the Fermi surface topology is drastically different for various FeHTS's, and likely plays a secondary role  in the superconductivity of FeHTS's. On the other hand, our new findings  of  the extraordinary  bandwidth-control and quasiparticle scattering properties of dopants in FeHTS's provide an alternative and likely unifying view angle to understand the complex phase diagrams of various series of FeHTS's, and their unconventional superconductivity.
For example,
the anomalous impurity scattering behaviors could explain (at least partially) the different residual electrical resistivities \cite{Nakajima}, the  robust superconductivity against heavy doping,  and the different maximal $T_C$'s and superconducting dome sizes in different series of FeHTS's.

The bandwidth-control of both heterovalent and isovalent dopants gives a natural explanation of their similar phase diagrams.  Moreover, we found that the increase of the bandwidth by doping is either in harmony with  the shrinking Fe-anion bond length, or the doped $4d$ transition metal [here it is ruthenium (Ru)] concentration, and the superconducting region corresponds to a quite ubiquitous bandwidth range. Therefore,  our finding  would help to bridge up the missing link between the  structural parameters and the electronic structures, that is, changing the lattice structure, such as bond length, will significantly alter the bandwidth and further affect the $T_C$.
Finally, our results  suggest that moderate bandwidth (or moderate correlation) plus minimal impurity scattering in the Fe-anion layer  are the essential factors for  maximizing $T_C$ in FeHTS's.

Therefore,  many puzzling and seemingly random phenomena of the FeHTS's could be   comprehended (at least a step forward) after realizing these multifold roles of doping. In particular, our results indicate that 
the bandwidth-control is most likely the primary control parameter  for FeHTS's rather than the filling-control, which  should be expected but unfortunately ignored so far, since the starting parent compound of FeHTS is a metal instead of a Mott insulator for cuprate superconductors.

\begin{figure*}[t]
\includegraphics[width=17.5cm]{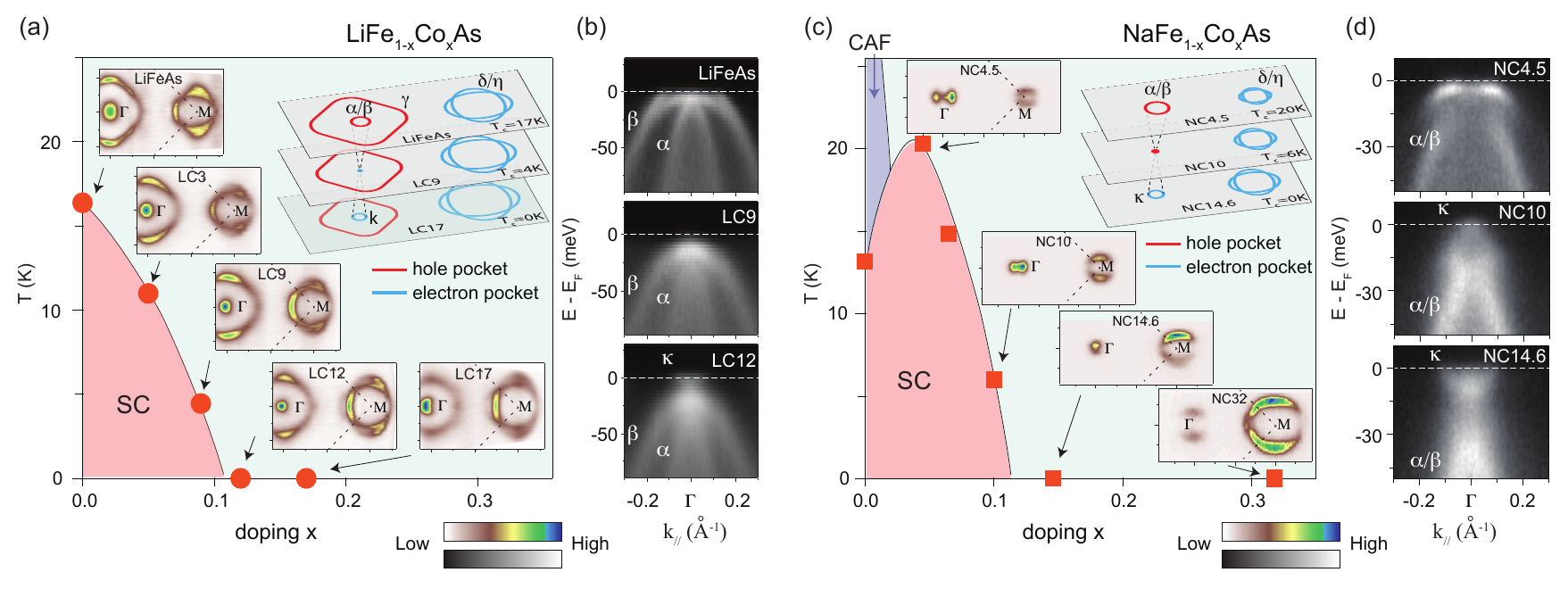}
\caption{(a) The phase diagram and corresponding photoemission intensity maps for LiFe$_{1-x}$Co$_x$As. The top-right inset panel illustrates the doping dependence of the Fermi surface topologies, where the hole and electron pockets are plotted with the red and blue lines, respectively.  (b) Doping dependence of the photoemission intensity distributions taken near the zone center along $\Gamma$~-~M direction for LiFe$_{1-x}$Co$_x$As. (c) and (d) are the same as panels (a) and (b), but for NaFe$_{1-x}$Co$_x$As. The superconducting and collinear antiferromagnetic phases are abbreviated as SC and CAF, respectively.}\label{fig1}
\end{figure*}


\section{Experiment}

Many FeHTS series were studied in this research, including two 111 series [NaFe$_{1-x}$Co$_{x}$As (ref.~\onlinecite{NaCophase}) and LiFe$_{1-x}$Co$_{x}$As (ref.~\onlinecite{LiCophase})],  three 122 series [Ba$_{1-x}$K$_x$Fe$_2$As$_2$ (ref.~\onlinecite{XHChenBaK}), BaFe$_2$(As$_{1-x}$P$_x$)$_2$ (ref.~\onlinecite{Zirong122}), and Ba(Fe$_{1-x}$Ru$_x$)$_2$As$_2$ (ref.~\onlinecite{BaRuphase})],  one 11 series [Fe$_{1.04}$Te$_{1-x}$Se$_x$ (ref.~\onlinecite{FChenFeTeSe})], and K$_x$Fe$_{2-y}$Se$_2$ (ref.~\onlinecite{XHChenKFeSe}) etc. For each series, high quality single crystals of various dopings were synthesized according to the cited references, which also give  corresponding phase diagrams. The samples are named by their dopant percentages throughout the paper. For example, the x~=~0, 0.03, 0.09, 0.12, 0.17, and 0.3 samples of LiFe$_{1-x}$Co$_{x}$As are named as LiFeAs, LC3, LC9, LC12, LC17, and LC30, respectively. ARPES measurements were performed at Fudan University with 21.2 eV light from a helium discharging lamp, and also at various beamlines, including the beamline 5-4 of Stanford Synchrotron Radiation Lightsource (SSRL), the beamline 1 and beamline 9A of Hiroshima Synchrotron Radiation Center (HiSOR) and the SIS beamline of Swiss Light Source (SLS). All the data were taken with Scienta R4000 electron analyzers. The overall energy resolution was 5$\thicksim$10 meV at Fudan, SSRL and HiSOR, or 15$\thicksim$20 meV at SLS depending on the photon energy, and the angular resolution was 0.3 degree. The samples were cleaved $\mathit{in~situ}$, and measured in ultrahigh vacuum with pressure better than 3$\times$10$^{-11}$ torr.

\begin{figure}[b]
\includegraphics[width=8cm]{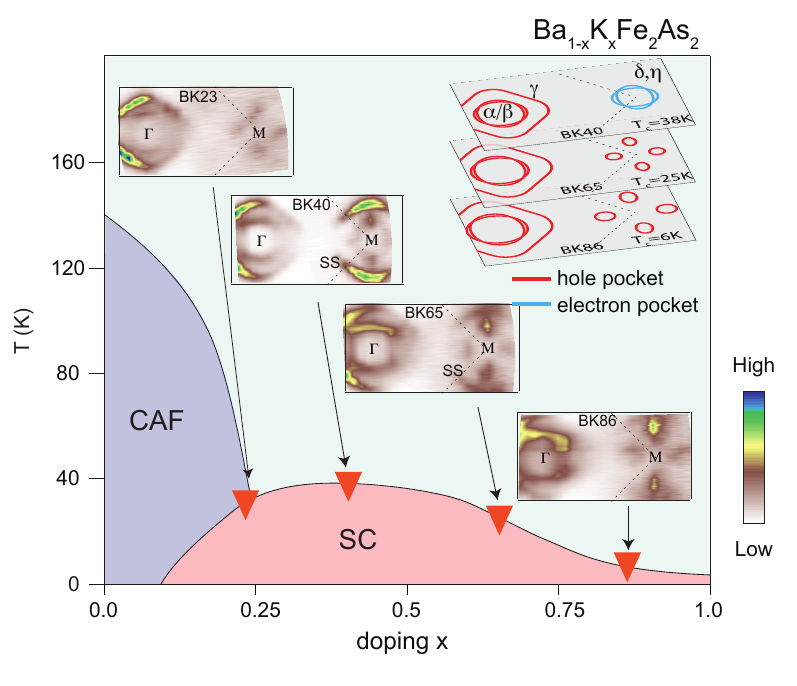}
\caption{The phase diagram and corresponding photoemission intensity maps for Ba$_{1-x}$K$_x$Fe$_2$As$_2$. The top-right inset panel illustrates the doping dependence of the Fermi surface topologies, where the hole and electron pockets are plotted with the red and blue lines, respectively. The phase diagram was extracted from ref.~\onlinecite{BaKphase}. Note that, the large pockets around the zone corner are the surface states (SS) due to the barium (Ba) reconstruction at the cleaved surface.}\label{fig2}
\end{figure}

\section{results}


\subsection{Filling-control: Fermi surface evolution and Lifshitz transition}

The Fermi surface in LiFeAs could represent the general Fermi surface topology of FeHTS's, 
which consists three hole pockets near the zone center and two electron pockets near the zone corner [Fig.~\ref{fig1}(a)].  
The inner hole pockets, $\alpha$ and $\beta$, are intertwined with each other and originate from the $d_{xz}$ and $d_{yz}$ orbitals, respectively. 
The outer $\gamma$ hole pocket is constructed by the $d_{xy}$ orbital. Around the zone corner, the  $d_{xz}$, $d_{yz}$, and $d_{xy}$ orbitals form the $\delta$ and $\eta$ electron pockets. For the heterovalent dopants, the obvious effect is to change the carrier density, with the sizes of the hole and electron Fermi pockets changing in opposite directions. 
As shown in Figs.~\ref{fig1}(a) and \ref{fig1}(c), with Co doping, the hole pockets shrink while the electron pockets enlarge, which indicates that replacing Fe with Co introduces electrons into the system. An opposite trend of Fermi surface evolution was observed for the hole-doped side,  where the hole pockets enlarge and the electron pockets shrink as shown in Fig.~\ref{fig2} for Ba$_{1-x}$K$_x$Fe$_2$As$_2$.

The Fermi surface topology would eventually  change with sufficient carrier doping. For the electron-doped case,  the band tops of the center hole bands shift downwards below the Fermi energy ($E_F$) with doping, and an electron band $\kappa$ could be observed in LC12 and NC14.6 [Figs.~\ref{fig1}(b) and \ref{fig1}(d)]. As a result, a Lifshitz transition occurs near the zone center for LiFe$_{1-x}$Co$_{x}$As and NaFe$_{1-x}$Co$_{x}$As. The $\alpha$ and $\beta$ hole pockets disappear and the $\kappa$ electron pocket emerges. Similar behavior was also observed in previous ARPES study on Ba(Fe$_{1-x}$Co$_x$)$_2$As$_2$ (ref.~\onlinecite{BaColifshitz}). The electron doping triggers a Lifshitz transition at the zone center while the hole doping could affect the topology of the Fermi pockets at the zone corner. As shown in Fig.~\ref{fig2}, in the heavily hole-doped  Ba$_{1-x}$K$_x$Fe$_2$As$_2$, the $\delta$ and $\eta$ electron pockets shift up above $E_F$  and four propeller-like hole pockets could be observed \cite{BaKlifshitz}.

As shown above, the sizes of the Fermi pockets can be tuned effectively by the carrier doping in FeHTS's, which alters the Fermi surface nesting condition and thus  affects the strength of low energy spin fluctuations as observed by nuclear magnetic resonance \cite{NMR1, NMR2, NMR3, NMR4, NMR5}.  However, the strength of such low energy spin fluctuations was found to be not sufficient to describe the superconductivity in FeHTS's \cite{Neutron2, Neutron1}. On the other hand, the change of the Fermi surface topology, or Lifshitz transition, has been proposed to be responsible for the disappearance of superconductivity or the pairing symmetry transition in the heavily doped compounds\cite{BaColifshitz, SplusD}. However, as will be argued later in the Discussions section, the Fermi surface topologies vary strongly in different series of  FeHTS's, which generally do not show a correlation with  the  superconductivity. This indicates that other effects of doping need to be considered.


\subsection{Quasiparticle scattering}

\begin{figure}[t]
\includegraphics[width=8.5cm]{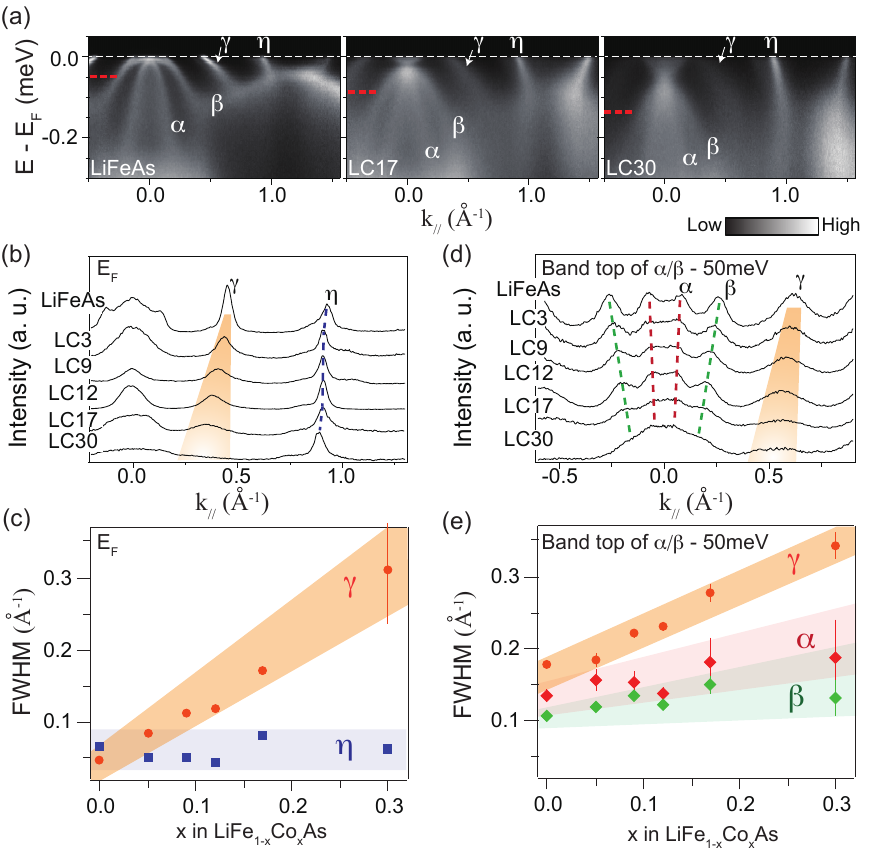}
\caption{(a) Doping dependence of the photoemission intensity distributions parallel to $\Gamma$~-~M direction in LiFe$_{1-x}$Co$_x$As, taken with mixed polarized photons. The red dashed lines illustrate the energy positions of the MDCs in panel (d). (b) Doping dependence of the MDCs at $E_F$.  (c) is the corresponding FWHMs of $\gamma$ and $\eta$ in panel (b).  (d)  MDCs at 50~meV below the band tops of $\alpha$ and $\beta$ as a function of doping, since $\alpha$ and $\beta$ do not cross $E_F$.  (e) is the corresponding FWHMs of $\alpha$, $\beta$, and $\gamma$ in panel (d). The error bars of FWHM in panels (c) and (e) are standard deviations of the Lorentzian fit to MDC peaks. }\label{impurity}
\end{figure}

\begin{figure}[t]
\includegraphics[width=8.5cm]{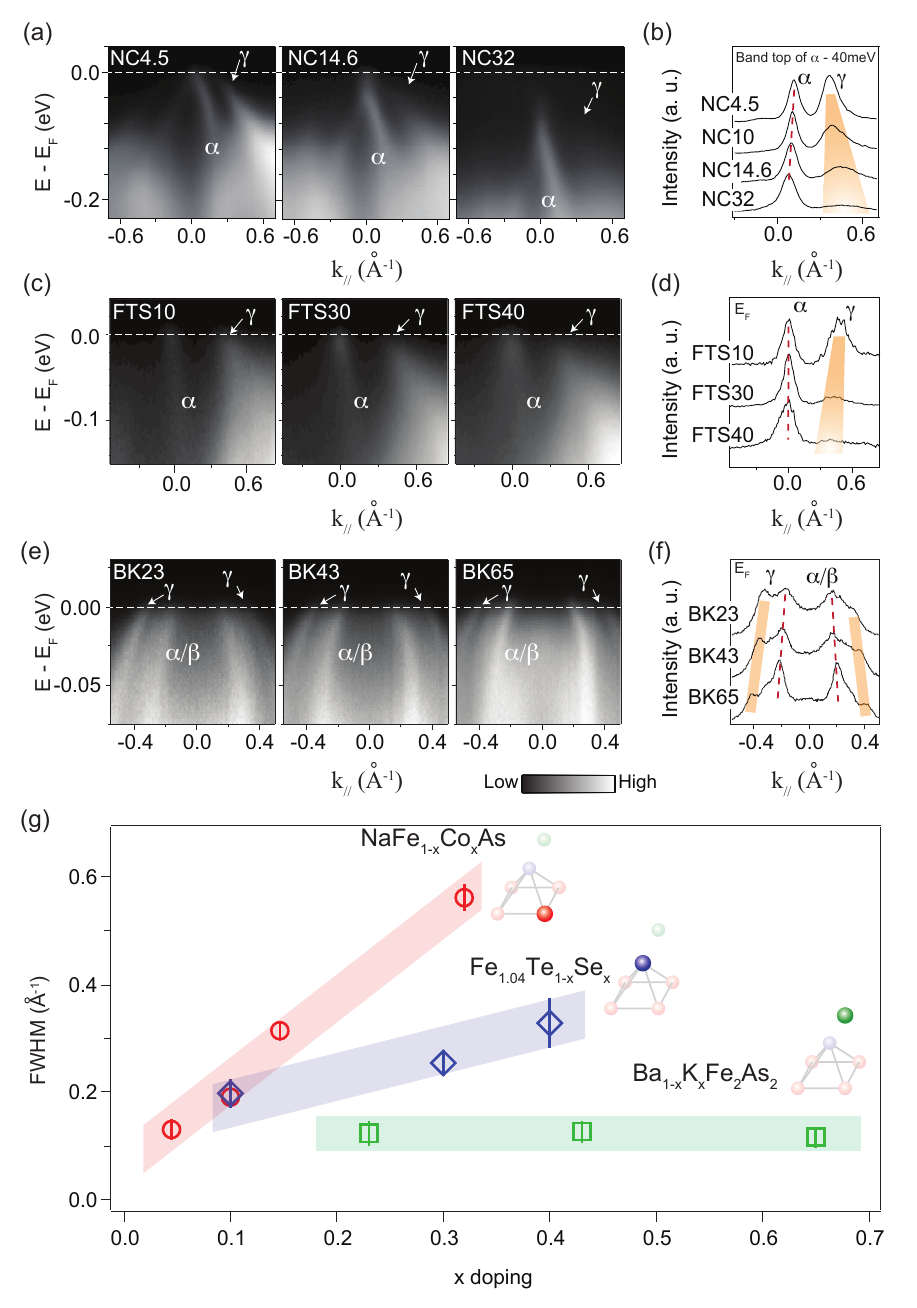}
\caption{(a) Doping dependence of the photoemission intensity distributions parallel to $\Gamma$~-~X direction in NaFe$_{1-x}$Co$_x$As. (b) Doping dependence of the MDCs at 50~meV below the band top of $\alpha$. Note that the increasing binding energy could only contribute a small increase on FWHM, which is less then 0.05$\AA^{-1}$ as shown in Fig.~\ref{impurity}(e).  (c) and (d), (e) and (f) are the same as (a) and (b), but for Fe$_{1.04}$Te$_{1-x}$Se$_x$ and Ba$_{1-x}$K$_x$Fe$_2$As$_2$, respectively. The MDCs in (d) and (f) were extracted from $E_F$. Note that, the data for NaFe$_{1-x}$Co$_x$As and Fe$_{1.04}$Te$_{1-x}$Se$_x$ were taken with $s$ polarized photons and those for Ba$_{1-x}$K$_x$Fe$_2$As$_2$ were taken with mixed polarized photons. (g) summarizes the doping evolutions of the FWHMs of the $\gamma$ bands in NaFe$_{1-x}$Co$_x$As, Fe$_{1.04}$Te$_{1-x}$Se$_x$ and Ba$_{1-x}$K$_x$Fe$_2$As$_2$ series.  The error bars of FWHMs in panel (g) are standard deviations of the Lorentzian fit to MDC peaks.}\label{impurity2}
\end{figure}

\begin{figure*}[t]
\includegraphics[width=15.4cm]{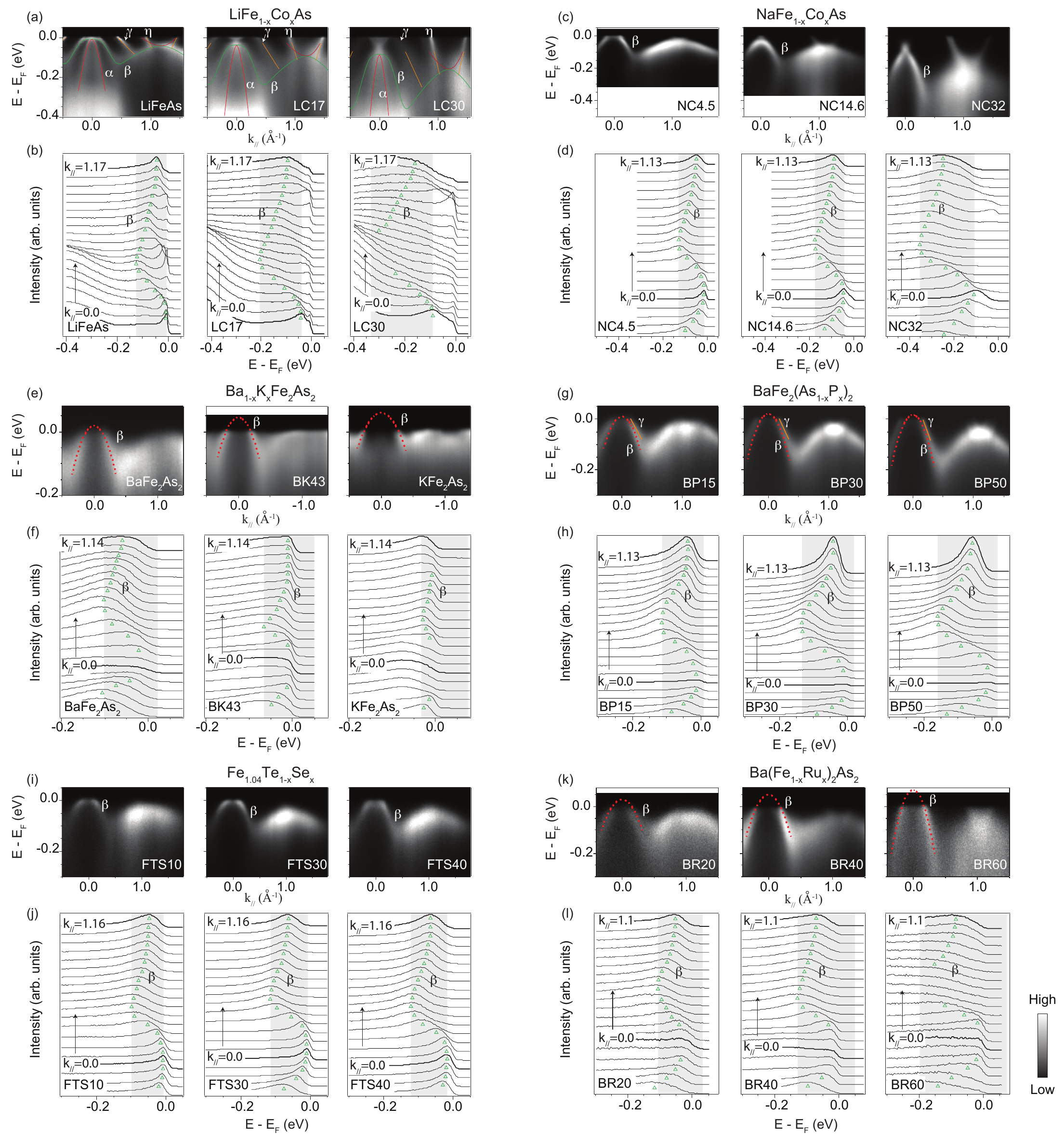}
\caption{(a) Doping dependence of the photoemission intensity distributions parallel to $\Gamma$~-~M direction in LiFe$_{1-x}$Co$_x$As, taken with mixed polarized photons. The band structure was determined based on the photoemission data in LiFeAs. The entire band structure was shifted in energy and normalized by a factor of $\sim 1.6$ in LC17 and $\sim 2.2$ in LC30. The obtained band structures were overlaid on the photoemission data in LC17 and LC30. (b) Corresponding energy distribution curves (EDCs) of the data in panel (a). The intensities of the EDCs were normalized to enhance the $\beta$ band. The green triangles trace the band dispersion of $\beta$. The gray shaded area is a guide to the eyes for viewing the change of the bandwidth with doping. (c), (e), (g), (i), and (k) Doping dependences of the photoemission intensity distributions taken parallel to $\Gamma$~-~M direction with $s$ polarized photons in NaFe$_{1-x}$Co$_x$As, Ba$_{1-x}$K$_x$Fe$_2$As$_2$,  BaFe$_2$(As$_{1-x}$P$_x$), Fe$_{1.04}$Te$_{1-x}$Se$_x$,  and Ba(Fe$_{1-x}$Ru$_x$)$_2$As$_2$, respectively. The photoemission data for each series are from the same $k_z$, although the bandwidth varies little with $k_z$ as reported before \cite{Zirong122}. Note that, the red dashed lines overlaid on panels (e), (g) and (k) are the quadratic curve fitting results for the $\beta$ bands, in order to determine the energy positions of $\beta$ band tops. (d), (f), (h), (j), and (l) are the same as (b), but for NaFe$_{1-x}$Co$_x$As,  Ba$_{1-x}$K$_x$Fe$_2$As$_2$, BaFe$_2$(As$_{1-x}$P$_x$), Fe$_{1.04}$Te$_{1-x}$Se$_x$, and Ba(Fe$_{1-x}$Ru$_x$)$_2$As$_2$, respectively.}\label{fig5}
\end{figure*}

\begin{figure*}[t]
\includegraphics[width=17cm]{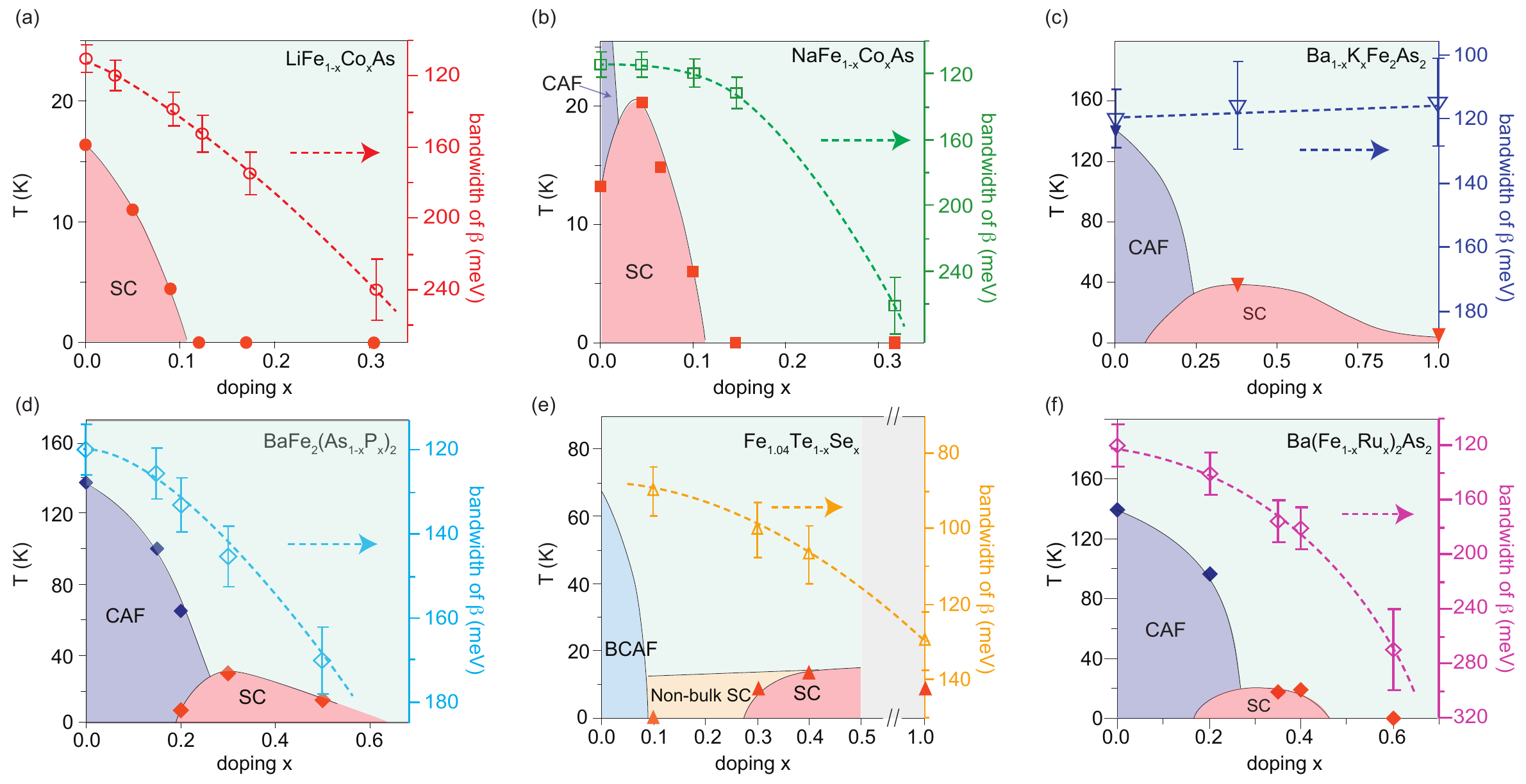}
\caption{(a) Evolutions of $T_C$ and  $\beta$ bandwidth with doping in  LiFe$_{1-x}$Co$_x$As. (b)~-~(f) are the same as (a), but for NaFe$_{1-x}$Co$_x$As, Ba$_{1-x}$K$_x$Fe$_2$As$_2$, BaFe$_2$(As$_{1-x}$P$_x$)$_2$, Fe$_{1.04}$Te$_{1-x}$Se$_x$, and Ba(Fe$_{1-x}$Ru$_x$)$_2$As$_2$, respectively. The bi-collinear antiferromagnetic phase is abbreviated as BCAF. The bandwidth for  Fe$_{1.04}$Se in panel (e) was extracted from ref.~\onlinecite{BorisenkoFeSe} and ref.~\onlinecite{TakahashiFeSe}. Note that, the doping range for 0.5$\textless$x$\textless$1.0 in Fe$_{1.04}$Te$_{1-x}$Se$_x$  can not be chemically synthesized \cite{FeTechemical}. The phase diagrams for Ba$_{1-x}$K$_x$Fe$_2$As$_2$, BaFe$_2$(As$_{1-x}$P$_x$)$_2$, Fe$_{1.04}$Te$_{1-x}$Se$_x$, and Ba(Fe$_{1-x}$Ru$_x$)$_2$As$_2$ were extracted from refs.~\onlinecite{BaKphase, Zirong122, FeTeSephase, BaRuphase}, respectively. The error bars of $\beta$ bandwidths  come from the  uncertainty in the dispersion determination. }\label{fig6}
\end{figure*}

\begin{figure}[t]
\includegraphics[width=8.5cm]{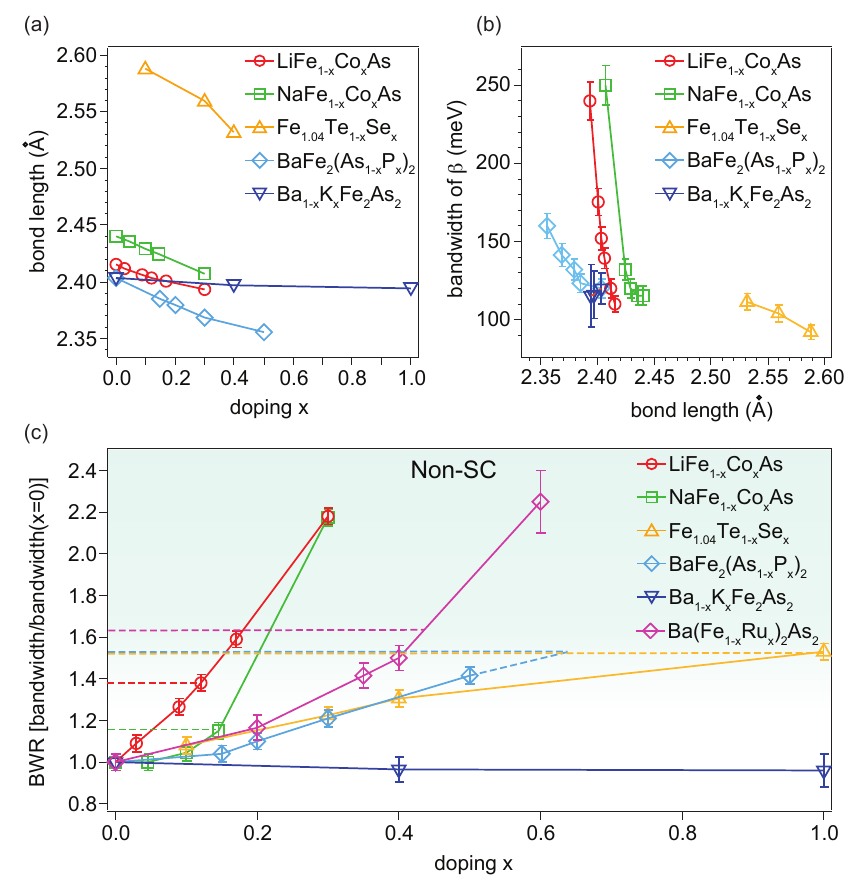}
\caption{(a) Doing evolutions of the Fe-As bond length in LiFe$_{1-x}$Co$_x$As, NaFe$_{1-x}$Co$_x$As, BaFe$_2$(As$_{1-x}$P$_x$)$_2$, Ba$_{1-x}$K$_x$Fe$_2$As$_2$,  or Fe-Te bond length in Fe$_{1.04}$Te$_{1-x}$Se$_x$. The bond length data were extracted from ref.~\onlinecite{bondlength}. Note that, in case of lacking the bond length data for certain doping level samples, we estimated the bond length values by the linear interpolation method. (b) summarizes the evolutions of the $\beta$ bandwidths as a function of the Fe-As or Fe-Te bond length in these five series. (c) Doping dependence of the  $\beta$ bandwidth normalized by that of its parent compound (x=0), named as BWR (bandwidth ratio) for simplicity, in each series. The horizontal dashed line divides the superconducting region and over-doped non-superconducting region of each series. Note that, the bandwidth of FeTe, parent compound of Fe$_{1.04}$Te$_{1-x}$Se$_x$, was estimated from a linear extrapolation of data in Fig.~\ref{fig6}(e), since the photoemission data of FeTe is intrinsically very broad \cite{YZhangFeTe}, which is hard to directly determine its bandwidth. The error bars of $\beta$ bandwidths or normalized bandwidths in panels (b) and (c) come from the uncertainty in the dispersion determination. }\label{correlation_sum}
\end{figure}

We now turn our focus to the impurity scattering effect, another effect that could be induced by dopants. As shown in Fig.~\ref{impurity}(a),  the $d_{xy}$-based $\gamma$ band becomes significantly weaker and broader with Co doping in LiFe$_{1-x}$Co$_x$As series. Figures~\ref{impurity}(b)~-~\ref{impurity}(e) plot the momentum distribution curves (MDCs) at $E_F$ and 50~meV below the band tops of the $\alpha$ and $\beta$ bands (so that the bands are resolvable), together with the full-width-half-maximum (FWHM) of each band, which reflects the scattering strength in each band. The FWHM of $\gamma$ increases remarkably with Co doping. This is not due to the increase of electronic correlations,  as the band renormalization decreases with the increase of dopants as will be shown in Fig.~\ref{fig6}(a) later. Therefore, the broadening of $\gamma$  observed here is more likely an impurity scattering effect. On the other hand, FWHM of  all the other bands do not change much with doping. We note there are
slight increases of FWHMs with doping for the $\alpha$ and $\beta$ bands  at 50~meV below their band tops
in Fig.~\ref{impurity}(e). However, this is  actually because the band tops of $\alpha$ and $\beta$ shift to higher  binding energies with increasing electron density, which would enhance the scatterings according to the Landau Fermi liquid theory. 
Therefore, the quasiparticle lifetimes of $\eta$, $\alpha$, and $\beta$ are essentially insensitive to the impurity scattering caused by the Co dopants, while the $\gamma$ band made of the $d_{xy}$ orbital is much more susceptible to the Co dopants. 
 
We also observed similar impurity scattering behavior in  NaFe$_{1-x}$Co$_x$As and Fe$_{1.04}$Te$_{1-x}$Se$_x$  [Figs.~\ref{impurity2}(a)~-~\ref{impurity2}(d)]. However in  Ba$_{1-x}$K$_x$Fe$_2$As$_2$, the impurity scattering to the quasiparticles appears to be absent [Figs.~\ref{impurity2}(e)~-~\ref{impurity2}(f)].
Figure~\ref{impurity2}(g) compares the doping dependences of the MDC FWHMs of the $\gamma$ bands as a function of doping for various series of compounds. The broadening of $\gamma$ is most pronounced in NaFe$_{1-x}$Co$_x$As and almost negligible in Ba$_{1-x}$K$_x$Fe$_2$As$_2$. The site dependence observed here further indicates that the broadening of $\gamma$ should originate from the impurity scattering induced by the dopants: when the dopant moves away from the Fe-anion layer, the scattering strength gradually decreases.


\subsection{Bandwidth-control}

Besides altering the Fermi surfaces and scattering the quasiparticles,  dopants also change the structure, thus change the band structure by changing electron hopping terms.
In BaFe$_2$(As$_{1-x}$P$_x$)$_2$, the bandwidth and Fermi velocity were found to increase significantly with P doping, indicating the decrease of  electronic correlations \cite{Zirong122}. Similar changes could also be observed for the Co-doped compounds. As shown in Fig.~\ref{fig5}(a), the band structure measured from  LiFeAs  could well match the bands in LC17 or  LC30, after it is  shifted in energy and scaled by a factor of  $\sim1.6$ or  $\sim 2.2$, respectively. This shows that the bandwidth increases equally for all the bands. Since only the  top and  bottom of the $\beta$ band  can be both observed in most cases, we take the bandwidth of $\beta$ as a characterization of the overall Fe $3d$ bandwidth [Fig.~\ref{fig5}(b)].

The same analysis on the  evolution of the $\beta$ bandwidth with doping  was extended to NaFe$_{1-x}$Co$_x$As, Ba$_{1-x}$K$_x$Fe$_2$As$_2$, BaFe$_2$(As$_{1-x}$P$_x$)$_2$,  Fe$_{1.04}$Te$_{1-x}$Se$_x$,  and Ba(Fe$_{1-x}$Ru$_x$)$_2$As$_2$, as shown in Figs.~\ref{fig5}(c)~-~\ref{fig5}(l). Note that, while the band top of $\beta$ is below $E_F$ in LiFe$_{1-x}$Co$_x$As and NaFe$_{1-x}$Co$_x$As [Figs.~\ref{fig5}(a) and \ref{fig5}(c)] or just touches $E_F$ in Fe$_{1.04}$Te$_{1-x}$Se$_x$ [Fig.~\ref{fig5}(i)], the $\beta$ bands in Ba$_{1-x}$K$_x$Fe$_2$As$_2$ [Fig.~\ref{fig5}(e)], BaFe$_2$(As$_{1-x}$P$_x$)$_2$ [Fig.~\ref{fig5}(g)], and Ba(Fe$_{1-x}$Ru$_x$)$_2$As$_2$ [Fig.~\ref{fig5}(k)] cross $E_F$ near the zone center. In order to determine the energy position of the band top of $\beta$, we applied a parabolic-curve fitting to the $\beta$ band dispersion  for every  doping. The fitted curves well follow the band dispersions of $\beta$ below $E_F$, and the fitted  effective mass of $\beta$ near the zone center 
shows  consistent doping dependence with the bandwidth in all three systems. We further quantitatively summarize the doping dependences of the bandwidths in various systems, as shown in Figs.~\ref{fig6}(a)~-~\ref{fig6}(f).  The increase of the bandwidth with doping is universal for all systems except for Ba$_{1-x}$K$_x$Fe$_2$As$_2$, where the bandwidth of $\beta$ shows a very small decrease, or is almost insensitive to the K doping after considering the error bars.  We will discuss the possible causes in detail later.

\section{discussions}

As we have shown above, the dopants could change the electronic structure in three different aspects: change the carrier concentration and alter the Fermi surface; scatter the quasiparticles of the central $d_{xy}$-based $\gamma$ band, whose strength strongly depends on the site of dopants; increase the bandwidths for various systems except for Ba$_{1-x}$K$_x$Fe$_2$As$_2$. In this section, we will discuss the implications of these findings, particularly on the superconductivity.

\subsection {Band-selective and site-dependent impurity scattering effects}

The dopants could significantly scatter the quasiparticles of the $d_{xy}$-originated $\gamma$ band around the zone center, while other bands are relatively unaffected. The scattering strength is the strongest when the dopant is in the Fe-anion layer. Such a band-selective and site-dependent impurity scattering effect needs further theoretical understandings. Nevertheless, our findings could explain many existing observations:

\begin{enumerate}

\item The superconductivity is  robust  against heavy doping in FeHTS's, since most bands are basically unaffected by the scattering of dopants.

\item It could partially explain why the maximal $T_C$'s of NaFe$_{1-x}$Co$_x$As and Fe$_{1.04}$Te$_{1-x}$Se$_x$ are lower than that of Ba$_{1-x}$K$_x$Fe$_2$As$_2$. Because the quasiparticle near $E_F$ is strongly suppressed for the large $\gamma$  Fermi pocket  in  NaFe$_{1-x}$Co$_x$As and Fe$_{1.04}$Te$_{1-x}$Se$_x$ [Figs.~\ref{impurity2}(a)~-~ \ref{impurity2}(d)], which thus likely does not  contribute to the superconductivity \cite{KurokiFS}.

\item  Similar to  Ba$_{1-x}$K$_x$Fe$_2$As$_2$, the superconductivity in the so-called 1111 series is  obtained by doping off the Fe-anion plane as well \cite{1111ZXZhao}.  The record high $T_C$ of 56~K in this series may be related to the weak scattering of the off-plane dopants.
Moreover,
the complete phase diagram of LaFeAsO$_{1-x}$H$_x$  exhibits a large  superconducting dome with  rather flat top, where its $T_C$ is independent of the doping \cite{1111Hosono}, similar to the case in Ba$_{1-x}$K$_x$Fe$_2$As$_2$. This shows that the impurity scattering strength caused by hydrogen (H), which is off the FeAs plane, should be weak in  LaFeAsO$_{1-x}$H$_x$ as well, and the superconductivity may be insensitive to carrier density variation over a large range.

\item It may partially explains  that doping range for the superconducting dome increases in the general order of compounds with Co dopants (typically very narrow),  those with P or Se dopants (typically covering a third to a half of the phase diagram), and those with off-plane K dopants (typically covering more than half the phase diagram).

\item It explains the  residual electrical resistivity decreases in the order of Co-doped, P-doped, and K-doped BaFe$_2$As$_2$ reported recently by ref. \onlinecite{Nakajima}.

\item  A recent STM study on NaFe$_{1-x}$Co$_x$As shows that the low energy electronic state is somehow insensitive to the Co dopants \cite{HHWen}.  Our results provide an explanation: the tunneling matrix element is dominated by the  $d_{xz}/d_{yz}$ states which extend out-of-plane and are inert to impurity scattering, rather than  the $\gamma$ band made of the in-plane $d_{xy}$ orbital.

\end{enumerate}

\subsection {Origin and critical role of  bandwidth-control}

For a correlated material, bandwidth is a critical parameter to characterize its itinerancy. The band renormalization factor, a ratio between the calculated bandwidth from density functional theory and the measured bandwidth, can be regarded as a measure of the correlation strength.
The ratio between the bandwidth and the relevant interaction term, such as on-site Coulomb repulsion, Hund's rule coupling, or exchange interactions, determines the properties of the material.

Intriguingly, the bandwidth is almost doping independent for the hole-doped compounds Ba$_{1-x}$K$_x$Fe$_2$As$_2$ [Fig.~\ref{fig6}(c)], while for  BaFe$_2$(As$_{1-x}$P$_x$)$_2$ and Fe$_{1.04}$Te$_{1-x}$Se$_x$,  the carrier density is almost unchanged but the bandwidth increases significantly [Figs.~\ref{fig6}(d) and \ref{fig6}(e)].  As shown in Fig.~\ref{correlation_sum}(a), the  bond length of Fe-As or Fe-Te decreases with the doping in LiFe$_{1-x}$Co$_x$As, NaFe$_{1-x}$Co$_x$As, BaFe$_2$(As$_{1-x}$P$_x$)$_2$, and Fe$_{1.04}$Te$_{1-x}$Se$_x$, because of the smaller ionic radii of the Co, P and Se dopants than those  of elements substituted by them.  In contrast, the K dopants in Ba$_{1-x}$K$_x$Fe$_2$As$_2$ are out of the FeAs plane and the bond length is thus unchanged with doping.  In Fig.~\ref{correlation_sum}(b), we plot the evolution of the $\beta$ bandwidth with the Fe-As or Fe-Te bond length in each series. One finds that the bandwidth of $\beta$ increases with the decrease of the bond length.  Therefore, the bandwidth evolution in FeHTS's is closely related to the change of structure parameters, such as bond length we found here.  Ba(Fe$_{1-x}$Ru$_x$)$_2$As$_2$ seems to be an exception, since its bond length increases slightly with doping \cite{BaRubondlength}. However, the large orbital radius of Ru $4d$ electron overcomes the enlarged bond length,  and thus enhances the bandwidth [Fig.~\ref{fig6}(f)].

 Intriguingly, the bond length shrinks in a similar rate with doping for the Co and P/Se dopants as shown in Fig.~\ref{correlation_sum}(a). However, the bandwidth increases much more significantly for Co-doped compounds [Fig.~\ref{correlation_sum}(c)]. Such an additional suppression of correlation could be attributed to  the enhanced screening effect induced by more carriers. Following this scenario, it is difficult to understand
the fact that the electronic correlation is not suppressed but rather slightly enhanced  in heavily hole-doped Ba$_{1-x}$K$_x$Fe$_2$As$_2$.
 To understand this dilemma, one has to realize that the parent compound of FeHTS is not a half-filled Mott insulator.
For a half-filling band system where electronic correlations originate from intra-band Coulomb interaction, such as cuprates, both hole and electron doping suppress the electronic correlation. The phase diagram is particle-hole symmetric. 
 FeHTS is a Fe $3d^6$ multi-band  system, as a result, it has been proposed that  the electronic correlations are  mainly due to  the Hund's rule coupling, $J_H$, instead of intra-band Coulomb interaction \cite{ZPYin}.
  In this case, the hole doping actually drives the system towards $3d^5$ state where the strength of Hund's interaction is strongest, which is likely counter-balanced by the screening effects, giving the observed doping-independent bandwidth.  On the other hand, the electron doping drives the system towards $3d^7$ state and further reduces the electronic correlation.
  Therefore, the particle-hole asymmetric bandwidth-control observed here could be viewed as a positive evidence for the importance of Hund's rule coupling in inducing the electronic correlations in FeHTS's.

For FeHTS's,  it was proposed that the superconductivity in FeHTS's could be mediated by  spin or orbital fluctuations \cite{Kuroki, Mazin, Kotani, GapReview}, while the strength of such  fluctuations is related to the electronic correlations. It has been numerically demonstrated that the system becomes superconducting only after the ratio between effective exchange interactions and bandwidth surpasses a certain value \cite{Seo}. The  effective  exchange interactions are roughly doping independent, as illustrated by fitting the spin waves measured in the neutron scattering experiments \cite{BaNiNeutron, Neutron1, Neutron2, Neutron3}.
Consistently, our data show that the system becomes non-superconducting in the over-doped regime, when the bandwidth is sufficiently large [Figs.~\ref{fig6}(a),~\ref{fig6}(b),~\ref{fig6}(d)~-~\ref{fig6}(f)], no matter whether it is due to enlarged bond length,  or due to   doping $4d$ electrons. The bandwidth thus seems to be a more universal control parameter than the structural parameters.
Quantitatively, the boundary in the  BWR (bandwidth ratio between the bandwidth and that of its parent compound) between the superconducting region and the over-doped non-superconducting region  is between $1.2$ and $1.6$ [Fig.~\ref{correlation_sum}(c)], depending on the series. In general, Co-doped series, LiFe$_{1-x}$Co$_x$As and  NaFe$_{1-x}$Co$_x$As here, have  smaller   boundary BWR values or narrower superconducting regions, which might be caused by the stronger impurity scatterings there.  Overall,  superconductivity can not be sustained for compounds with BWR above $\sim 1.5$  in the FeHTS's  studied here. Taking the end members of BaFe$_2$P$_2$ and LaOFeP for examples, the quantum fluctuations in these two compounds are strongly suppressed by P dopants and the two systems  were reported to behave more like  normal metals with  large bandwidths \cite{BaFeP, LaOFeP}.
On the other hand, when the bandwidth is too small, or the correlation is too strong,
 the system is in the magnetic or orbital ordered  phase  [Figs.~\ref{fig6}(b)~-~\ref{fig6}(e)], and the competing order would suppress superconductivity.
For example in FeTe, the normal state shows semiconductor behavior, and the magnetic moment is as large as 2$\mu_B$ in the low-temperature magnetic ordered states \cite{FeTestrong1,FeTestrong2}.
Therefore,
our results thus suggest that the superconductivity in FeHTS's is  optimized at the moderate bandwidth, and  provide an explanation on  the fact that the phase diagrams of heterovalent doping and isovalent doping cases are similar.
Furthermore, the sensitivity of the electronic structure on the structural parameters demonstrated here  partially establishes the connection between  $T_C$ and the structural parameters.

\subsection {The secondary role of filling-control on superconductivity}

The Fermi surface topology was considered to be a dominating factor in FeHTS's. However, many debates and contradictions have been raised recently and the central question is whether the inter-pocket scattering between hole and electron pockets is critical for the superconductivity or not.
The correlation between the vanishing superconductivity and  the Lifshitz transition of the  hole pocket in Ba(Fe$_{1-x}$Co$_x$)$_2$As$_2$ reported before \cite{BaColifshitz}, and  in NaFe$_{1-x}$Co$_{x}$As and  LiFe$_{1-x}$Co$_x$As observed here, can be viewed as  support for the possible crucial role of the inter-pocket scattering between the central hole pockets and the corner electron pockets on the superconductivity \cite{Kuroki, Mazin}.
We note that although the $d_{xy}$-based hole pocket is present even in the heavily doped compounds LC17 and LC30, the quasiparticle near $E_F$ is ill-defined due to the strong impurity scattering and cannot contribute to any superconducting pairing.
However, such a picture has been seriously challenged by the recent studies on K$_x$Fe$_{2-y}$Se$_2$ and the monolayer FeSe thin film on a SrTiO$_3$ (STO) substrate, where the $T_C$'s are above 30$~$K,  but the Fermi surfaces are composed of only electron pockets without any central hole pocket \cite{YZhangKFeSe, DLFeSe, XJFeSe}.
 Scattering between the electron pockets around the zone corner was suggested to be sufficient for superconductivity in these iron selenides \cite{dwave}.
 One explanation is that the superconducting mechanisms of these iron selenides are remarkably different from the other FeHTS's. Other factors should be considered, for example, the phase separation between superconducting and insulating phases in K$_x$Fe$_{2-y}$Se$_2$ (refs.~\onlinecite{FChenPRX, LiWeiKFS}) and the critical role of substrate and interface in ultra-thin FeSe film \cite{DLFeSe2, DHLee}.
 
 \begin{figure}[t]
\includegraphics[width=8.5cm]{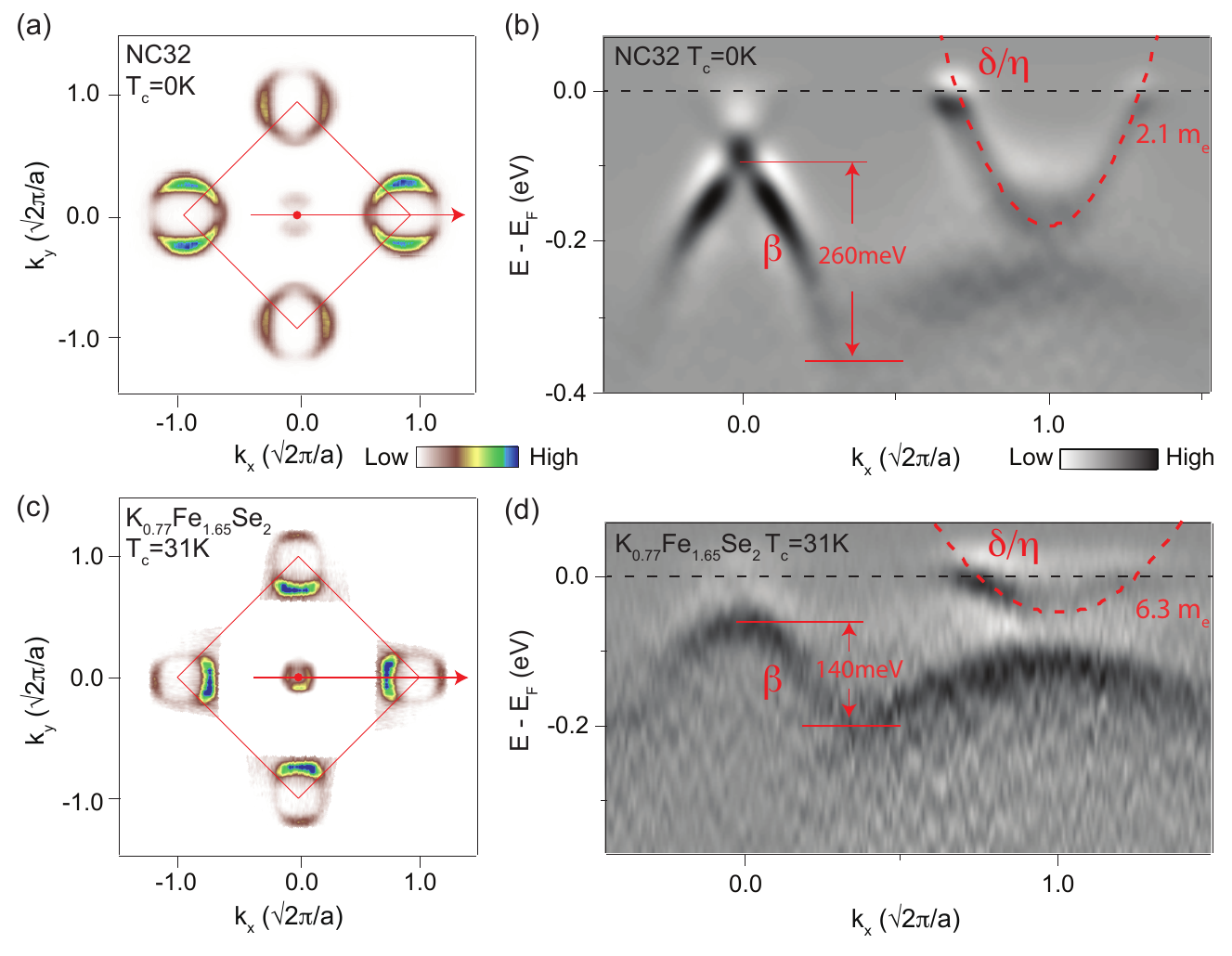}
\caption{(a) Photoemission intensity map across the $Z$ point for NC32, taken with 100 eV photons in $s$ polarization. (b) The photoemission intensity distribution  along $Z$~-~$A$ direction as illustrated by the red arrow in panel (a) for NC32, taken with 100 eV photons in $s$ polarization. (c) Photoemission intensity map across the $Z$ point for K$_{0.77}$Fe$_{1.65}$Se$_2$, taken with 31 eV photons in mixed polarization. (d) The photoemission intensity distribution along $Z$~-~$A$ direction for K$_{0.77}$Fe$_{1.65}$Se$_2$, taken with 121 eV photons in $s$ polarization. Note that, both  31 eV and 121 eV photons correspond to the $Z$ point in the Brillouin zone for K$_{0.77}$Fe$_{1.65}$Se$_2$.}\label{fig7}
\end{figure}

\begin{figure}[t]
\includegraphics[width=8.5cm]{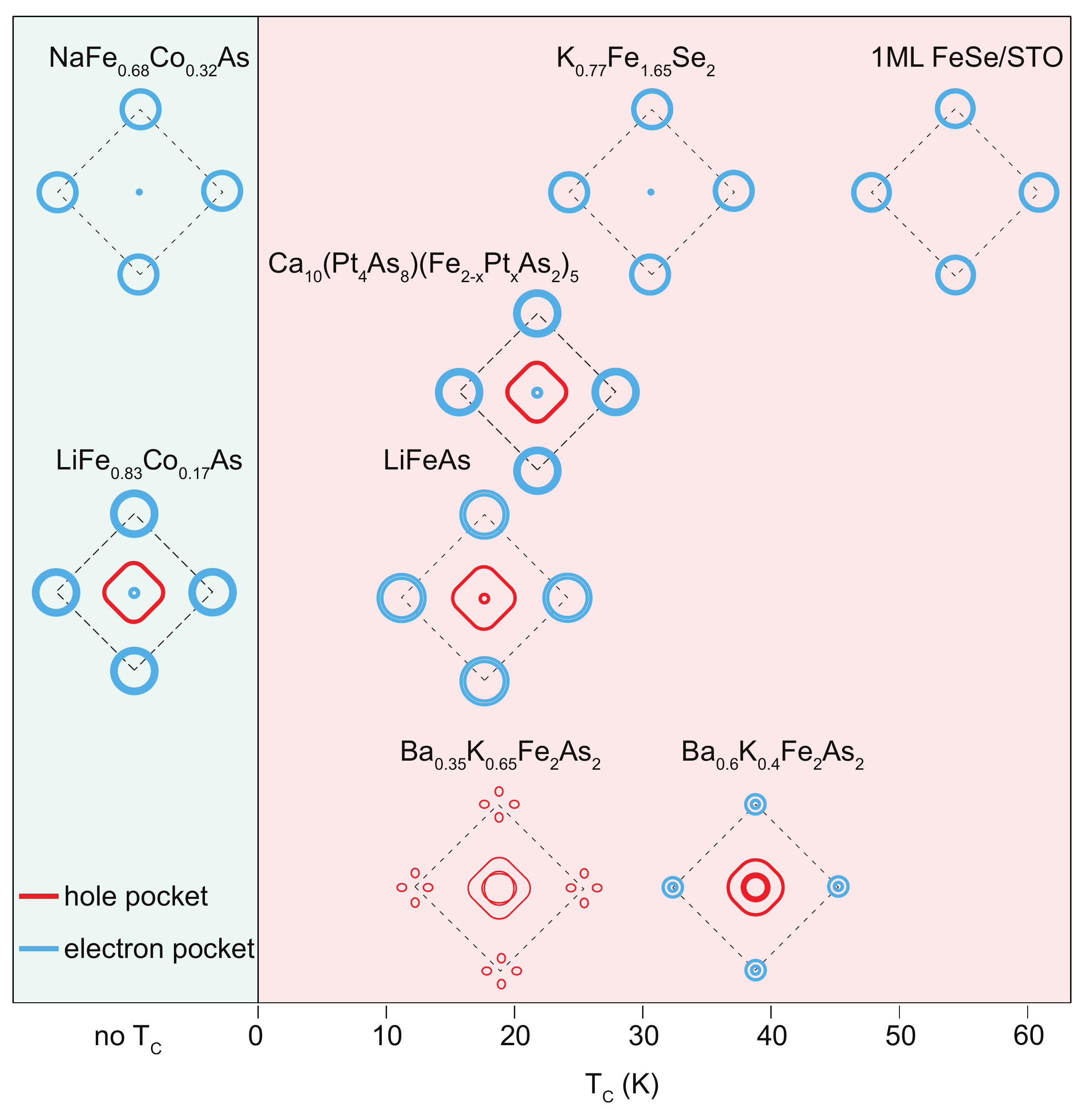}
\caption{Summary of the relation between Fermi surface topology and $T_C$ for different compounds in FeHTS's. All the Fermi surfaces were taken across the $\Gamma$ point. The hole pockets and electron pockets are illustrated with red and blue lines, respectively. $T_C$ is not directly correlated with the Fermi surface topology. The Fermi surfaces of K$_{0.77}$Fe$_{1.65}$Se$_2$, 1ML FeSe/STO, and Ca$_{10}$(Pt$_4$As$_8$)(Fe$_{2-x}$Pt$_x$As$_2$)$_5$ were extracted from refs.~\onlinecite{YZhangKFeSe, DLFeSe, CaFePtAs}, respectively.}\label{mapsum}
\end{figure}

If the superconductivities in iron-pnictides and iron-selenides share  a unified mechanism, the correlation between the Lifshitz transition and superconductivity observed in the Co-doped systems could be accidental. This is because, with the increasing  Co concentration, not only the Fermi surface topology is changed,  the electronic correlation also decreases at the same time [Figs.~\ref{fig6}(a) and \ref{fig6}(b)], which could strongly suppress the pairing strength for superconductivity. It is intriguing to compare the electronic structure of NC32 with that of K$_{0.77}$Fe$_{1.65}$Se$_2$, since NC32 and K$_{0.77}$Fe$_{1.65}$Se$_2$ possess a similar Fermi surface topology and size, but one is non-superconducting and the other has a $T_C$ above 30~K [Figs.~\ref{fig7}(a) and \ref{fig7}(c)].  Figures \ref{fig7}(b) and \ref{fig7}(d) compare their low-lying band structures. The difference is obvious. The larger bandwidth of the $\beta$ band and  the smaller effective mass of the $\delta/\eta$ electron band in NC32 than those in K$_{0.77}$Fe$_{1.65}$Se$_2$
all indicate that the electronic correlation in NC32 is much weaker than that of K$_{0.77}$Fe$_{1.65}$Se$_2$. If we compare the electronic structure of NC32 with the band calculation of NaFeAs after a rigid band shift, we could get a renormalization factor of $\sim 1.8$ for NC32, which is smaller than both the factor of  $\sim 4$ in NaFeAs, and the factor of $\sim 3$ in K$_{0.77}$Fe$_{1.65}$Se$_2$ (refs.~\onlinecite{NaFeAsLDA,KFeSeLDA}). 
This also implies that the bandwidth of   K$_{0.77}$Fe$_{1.65}$Se$_2$ is moderately renormalized, and the superconductivity in these compounds  whose Fermi surfaces consist of only electron pockets is 
consistent with the picture presented in Fig.~\ref{correlation_sum}(c) as well.

The comparison between NC32 and K$_{0.77}$Fe$_{1.65}$Se$_2$  proves that  the same Fermi surface topology could  give dramatically different $T_C$'s. Another similar example is LC17 and  Ca$_{10}$(Pt$_4$As$_8$)(Fe$_{2-x}$Pt$_x$As$_2$)$_5$ - they have a similar  Fermi surface topology 
 with a $d_{xy}$-based hole pocket and an electron pocket around the zone center (Fig.~\ref{mapsum}) \cite{CaFePtAs}, however the $T_C$ is 22~K for Ca$_{10}$(Pt$_4$As$_8$)(Fe$_{2-x}$Pt$_x$As$_2$)$_5$ while 0 K for LC17. 
These comparisons  indicate that superconductivity does not rely on the presence of the  $d_{xz}/d_{yz}$-based hole pocket around the zone center, or even the presence of hole pocket at all.
On the other hand,  completely different Fermi surface topologies can host superconductivity of a similar strength.  As shown in Fig.~\ref{mapsum},   the superconductivity could emerge on Fermi surface  consisting of only electron pockets (K$_{0.77}$Fe$_{1.65}$Se$_2$ and 1ML FeSe/STO), only hole pockets (Ba$_{0.35}$K$_{0.65}$Fe$_2$As$_2$), both hole and electron pockets (LiFeAs and Ba$_{0.6}$K$_{0.4}$Fe$_2$As$_2$ et al.) or with some special Fermi surface forms [such as Ca$_{10}$(Pt$_4$As$_8$)(Fe$_{2-x}$Pt$_x$As$_2$)$_5$ with both hole and electron pockets around the zone center].

Considering all the facts shown above, we conclude  that the Fermi surface topology may just play a secondary role in determining $T_C$.  Other factors, such as the bandwidth (or relatedly,  correlation strength) and impurity scattering discussed in the last two subsections could play  more important roles. We also note that, when the impurity scattering strength and the  bandwidth are both less sensitive to the dopants, as the case in Ba$_{1-x}$K$_x$Fe$_2$As$_2$ [Figs.~\ref{impurity2}(g) and \ref{correlation_sum}(c)], the Fermi surface might play the leading role in determining $T_C$. As shown in Fig.~\ref{fig6}, for Ba$_{1-x}$K$_x$Fe$_2$As$_2$, the $T_C$ decreases much more slowly in the over-doped regime of the phase diagram compared with the other  systems. The suppression of $T_C$ in Ba$_{1-x}$K$_x$Fe$_2$As$_2$ was proposed to be due to the competition between the $s$-wave and $d$-wave pairing channels in the heavily doped compounds \cite{SplusD}, whose  strengths depend on the Fermi surface topology.

\section{Conclusions}

To summarize, out of the diversified materials and electronic structures of various series of FeHTS's, we have uncovered a unifying theme of the doping effects: the bandwidth-control by both heterovalent and isovalent dopants, and the band-selective and site-dependent impurity scattering effects, for the first time. Together with the usual filling-control, these provide a microscopic and comprehensive understanding of chemical substitution in FeHTS's.

Particularly,  we identified the most likely dominating role of the bandwidth (or equivalently, electronic correlation) on the superconductivity in FeHTS's, which provides a natural understanding of the similar phase diagrams obtained by various dopants. We further demonstrated that the bandwidth-control is closely related with the structure parameters, such as bond length. The different scattering effects and different structures may affect the maximal value of $T_C$, and cause the superficial diversity and complexity.
On the other hand, Fermi surface topology and its evolution with doping may play a secondary role in determining $T_C$.

The implications of our experimental findings  are profound and many-fold. It explains many puzzles and controversies, and unifies our current understanding on the phase diagrams, resistivity behaviors, superconducting properties etc.
Our data also suggest that one  need to minimize the impurity scattering in the Fe-anion layer while optimizing a moderate bandwidth in order to  enhance the $T_C$ record in the search of new FeHTS's.
Furthermore, these results put strong constrains on the  theories of the superconducting mechanism in FeHTS's. As the $T_C$ is less sensitive to the  Fermi surface topology, the weak-coupling theoretical scenario driven by the Fermi surface  should be reexamined \cite{Kuroki, Mazin}. Alternatively, the strong-coupling pairing scenario, where the superconducting pairing is mediated by the local antiferromagnetic exchange interaction\cite{JPHu}, is favored, since the exchange interaction is sensitive to the change of electronic correlation.

\section{Acknowledgements}

We gratefully acknowledge the experimental support by Dr. D. H. Lu, Dr. H. Makoto at SSRL, Dr. M. Shi, Dr N. Plumb at SLS, and Prof. K. Shimada, Dr. M. Arita, Dr. J. Jiang at HiSOR, and the helpful discussions with Prof. Jiangping Hu, and Prof. Zhongyi Lu. This work is supported in part by the National Science Foundation of China, and National Basic Research Program of China (973 Program) under the grant Nos. 2012CB921402, 2011CBA00112, 2011CB921802. SSRL is operated by the US DOE, Office of BES, Divisions of Chemical Sciences and Material Sciences.

\end{document}